\documentclass[11pt]{article}
\usepackage[utf8]{inputenc}

\newcommand{\linkeddata}{Linked Data}
\newcommand{\ld}{LD}
\newcommand{\dpr}{DP}
\newcommand{\ai}{AI}
\newcommand{\dmml}{DM/ML}
\newcommand{\datamining}{Data Mining}
\newcommand{\dmccschema}{\textit{dmcc-schema}}
\newcommand{\cc}{CC}

\newcommand{\sparql}{\textit{SparQL}}
\newcommand{\rdf}{RDF}
\newcommand{\sla}{SLA}
\newcommand{\dm}{DM}
\newcommand{\rf}{RF}
\newcommand{\rlang}{R}
\newcommand{\ml}{ML}
\newcommand{\turtle}{\textit{Turtle}}
\newcommand{\sw}{Semantic Web}
\newcommand{\mup}{Month Uptime Percentage (MUP)}
\newcommand{\azure}{Azure}
\newcommand{\apikey}{API KEY}
\newcommand{\restfulapi}{RESTful API}
\newcommand{\amazon}{Amazon}
\newcommand{\ram}{RAM}
\newcommand{\cpu}{CPU}

\newcommand{\is}{Internet of  Services}
\newcommand{\soa}{SoA}
\newcommand{\wsdl}{WSDL}
\newcommand{\sawsdl}{SA-WSDL}
\newcommand{\msm}{MSM}
\newcommand{\soaml}{SoAML}
\newcommand{\uddi}{UDDI}
\newcommand{\uml}{UML}
\newcommand{\xml}{XML}
\newcommand{\wadl}{WADL}
\newcommand{\occml}{OCCML}
\newcommand{\weka}{Weka}
\newcommand{\knime}{Knime}
\newcommand{\owls}{OWL-S}
\newcommand{\wsmo}{WSMO}
\newcommand{\sarest}{SA-REST}
\newcommand{\wsmolite}{WSMO-lite}
\newcommand{\linkedusdl}{Linked-USDL}
\newcommand{\usdl}{USDL}
\newcommand{\slo}{SLO}

\usepackage{booktabs}
\usepackage{hyperref}
\usepackage{xcolor}
\usepackage{float}
\usepackage{listings}
\usepackage{multirow}
\usepackage{listings}
\usepackage{epsfig}
\usepackage{xspace}
\usepackage{amsmath}
\usepackage{url}

\usepackage[left=2cm, right=2cm, top=2cm, bottom=2cm]{geometry}

\title{Semantics of Data Mining Services in Cloud Computing}
\author{Manuel~Parra-Royon\footnote{Department of Computer Sciences and Artificial Intelligence, Soft Computing and Intelligent Systems group, University of Granada, Spain.}, 
        ~Ghislain~Atemezing \footnote{Department of R\&D, Mondeca, Paris, France.}~and~Jose Manuel Ben\'itez-Sanchez\footnote{Department of Computer Sciences and Artificial Intelligence, Soft Computing and Intelligent Systems group, University of Granada, Spain.}}

\begin{document}

\maketitle

    \begin{abstract}
     
In recent years with the rise of Cloud Computing (\cc{}), many companies providing services in the cloud, are empowering a new series of services to their catalogue, such as data mining (\dm{}) and data processing (\dpr{}), taking advantage of the vast computing resources available to them. Different service definition proposals have been put forward to address the problem of describing services in \cc{} in a comprehensive  way. Bearing in mind that each provider has its own definition of the logic of its services, and specifically of \dm{} services, it should be pointed out that the possibility of describing services in a flexible way between providers is fundamental in order to maintain the usability and portability of this type of \cc{}  services. The use of semantic technologies based on the proposal offered by Linked Data (\ld{}) for the definition of services, allows the design and modelling of \dm{} services, achieving a high degree of interoperability. In this article a schema for the definition of \dm{} services on \cc{} is presented considering all key aspects of service in \cc{}, such as prices, interfaces, Software Level Agreement (\sla{}), instances or  \dm{} workflow, among others. The new schema  is based on \ld{}, and it reuses other schemata obtaining a better and more complete definition of the services. In order to validate the completeness of the scheme, a series of \dm{} services have been created where a set of algorithms such as \textit{Random Forest} (\rf{}) or \textit{KMeans} are modeled as services. In addition, a dataset has been generated including the definition of the services of several actual CC DM providers, confirming the effectiveness of the schema.

    \end{abstract}

\section{Introduction}
\label{sec:introduction}

Cloud Computing has been introduced into our daily lives in a completely transparent and friction-less way. The ease of Internet access and the exponential increase in the number of connected devices has made it even more popular. Adopting the phenomenon of \cc{} means a fundamental change in the way Information Technology (IT) services are explored, consumed or deployed. \cc{} is a model of providing services to companies, entities and users, following the utility model, such as energy or  gas. \cc{} can be seen as a model of service provision where computer resources and computing power are contracted through the Internet of services (IS)\cite{liu2016services}. A big part of the \cc{} services providers are currently leveraging their wide  computing infrastructure to offer a set of web services to enterprises, organizations and users. 

The increase in the volume of data generated by companies and organizations is growing at an extremely high rate. According to Forbes \cite{bernardmarr2016}, in 2020, the growth is expected to continue and data generation is predicted to increase by up to 4,300\%, all motivated by the large amount of data generated by service users. By 2020, it is estimated that more than 25 billion devices will be connected to the Internet, according to \textit{Gartner} \cite{gartner26b}, and that they will produce more than 44 billion GB of data annually. In this scenario, \cc{} providers have successfully included data analysis services in their catalogue of services for massive processing of data and \dm{}. 


These services allow the application of Artificial Intelligence (\ai{}) and Machine Learning (\ml{}) techniques on a large variety of data, offering an extensive catalogue of algorithms and workflows related to \dm{}. Services, such as \textit{Amazon SageMaker}\footnote{\url{https://aws.amazon.com/sagemaker/}} or   \textit{Microsoft Azure Machine Learning Studio} \footnote{\url{https://azure.microsoft.com/en-us/services/machine-learning-studio/}}  (Table \ref{dmservices}), offer a set of algorithms as services within \cc{}  platforms. Following this line, other \cc{} platforms such as \textit{Algorithmia}\footnote{\url{https://algorithmia.com/}} or \textit{Google Cloud ML}\footnote{\url{https://cloud.google.com/products/machine-learning/}}, offer \ml{} services at the highest level, providing specific services for the detection of objects in photographs and video, sentiment analysis, text mining or forecasting, for instance. 

Each \cc{} service provider offers a narrow definition of these services, which is generally incompatible with other service providers. For instance, where one provider has a service for \rf{} algorithm, another provider has another name, features, or parameters for that algorithm, although the two might be the same. This makes it difficult to define services or service models independent of the provider as well as to compare services through a \cc{} service broker \cite{lin2016cloud}. Indeed, a standardization of the definition of services would boost competitiveness, allowing third parties to operate with these services in a totally transparent way, skipping the individual details of the  providers. The effectiveness of \cc{} would be greatly improved if there were a general  standard for services definition \cite{ghazouani2017survey}.

There are several proposals for the definition of services. These proposals cover an important variety of both syntactic and semantic languages in order to achieve a correct definition and modelling of services. For the definition of this type of \dm{} services, there is no specific proposal, due to the complexity of the services represented. Solutions based on the proposal offered by \linkeddata{} \cite{bizer2009linked} can solve the problem of defining services from a perspective more comprehensive. \linkeddata{} undertakes models and structures from the \sw{}, a technology that aims to expose data on the web in a more reusable and inter-operable way with other applications.  The \linkeddata{} proposal allows you to link data and concept definitions from multiple domains through the use of the \sw{} \cite{berners2001semantic} articulated with  \textit{\rdf{}} \cite{klyne2006resource} or  \turtle{} \cite{beckett2008turtle} languages.

The main objective of this work is the definition of \dm{} services for \cc{} platforms  taking into account the \linkeddata{} principles. The definition of the service is not only focused on the main part of the service (algorithms, workflow, parameters or models), but also allows the definition and modelling of prices, authentication, \sla{}, computing resources or catalogue. The \dmccschema{} proposal provides a complete vocabulary for the exchange, consumption and negotiation of \dm{} services in \cc{}.  With this schema it is possible to make queries in \textit{SparQL} \cite{prud2006sparql} about this type of \cc{} \dm{} services and obtain, for example, the set of providers that offer a certain algorithm, as well as the economic cost of the service. This allows for the comparison between different providers and \dm{} services from varying points of view (costs, regions, instances, algorithms, etc). For the modelling of the different components of the service, existing schemata and vocabularies have been used, which have been adapted to the problem of the definition of \dm{} services. In addition, new vocabularies have been created to cover specific elements of the domain that have not been available up until now. We present \dmccschema{}, a proposal based on \linkeddata{} to cover the entire definition of \dmml{} cloud services and that allows the exchange, search and integration of this type of services in \cc{}.

The paper is structured as follows: In the following section we present the related works in \cc{} services definition. Section \ref{sec:dmcc-schema} put forward our proposal, the \dmccschema{} in detail, and depicts all components with their interactions. We describe in section \ref{sec:usecases} some use cases for a real-world data mining service, providing with a \rf{} algorithm as service and including some aspects related to the service definition in \cc{}, such as \sla{} or prices of the service, and the algorithm description. Finally, we conclude our work in section \ref{sec:conclusions}.

\section{Related work}
\label{sec:relatedwork}

\dm{} and \ml{} is a very highly relevant topic nowadays. These areas of knowledge provides entities, organizations and individuals with tools for data analysis and knowledge extraction. With the growth of \cc{} and Edge Computing \cite{shi2016edge}, \dm{} services  are taking a significant position in the catalogue offered by providers. The complexity of \dmml{}  services requires a complete specification, that is not limited to technical or conceptual considerations. The definition, therefore, must integrate key aspects of the service into the \cc{} environment.

The definition of services in general has been approached from multiple perspectives. At the syntactic level with the reference of Services-Oriented-Architecture (\soa{}) \cite{newcomer2005understanding} and \xml{} (Extended Mark-Up Language)  \cite{bray1997extensible}, derived languages such as \wsdl{} \cite{w3cwsdl}, \wadl{} \cite{hadley2006web}, or \soaml{} \cite{newcomer2005understanding}, \uddi{}  \cite{bellwood2002uddi} (both related to  \uml{} \cite{dennis2015systems}), it has been attempted to specify at the technical level the modelling of any \is{}.

The definition of services through semantic languages, enables to work with the modelling of services in a more flexible mode. Semantic languages allow to capture functional and non-functional features. \owls{} \cite{martin2004owl} integrates key factors for these services such as service discovery, process modelling, and service details. Web services can be described by using \wsmo{} \cite{domingue2005web} as well. With \wsmo{} it is possible to describe semantically the discovery, invocation and logic aspects of the service. Schemata such as \sawsdl{} \cite {kopecky2007sawsdl}, which integrate both semantic and syntactic schema, associate semantic annotations to \wsdl{} elements. \sarest{} \cite{klusch2014service}, \wsmolite{} \cite{kopecky2009hrests} or Minimal Service Model \cite{taheriyan2012rapidly} (\msm{}), are also part of the group for the exchange, automation and composition of services, focused on the final service. On the other hand \usdl{} \cite{kona2009usdl} considers  a global approach to service modelling that seeks to emphasize business, entity, price, legal, composition, and technical considerations. With \linkedusdl{}  \cite{pedrinaci2014linked}, this idea becomes concrete, in services integrated into the \cc{} model. \linkedusdl{} assumes an important part of the key aspects that a service must provide, extending its usefulness to the ability to create \cc{} services from scratch. With \linkedusdl{}, the modelling of entities, catalogue, \sla{} or interaction are considered.

The proposals address the modelling and definition of services in a generic mode, without dealing with the specific details of a \dmml{} services in \cc{}. For the treatment of these problems we typically use software platforms such as \weka{} \cite{Hall:2009:WDM:1656274.1656278}, \knime{} \cite{BCDG+07} or \textit{Orange} \cite{JMLR:demsar13a} and frameworks and libraries integrated into the most modern and efficient programming languages such as \textit{C}, \textit{Java}, \textit{Python}, \textit{R}, \textit{Scala} and others.  These environments lack elements for  \cc{} services modelling due to its nature. 


There are several approaches to tackle an ontology-based  languages for \dmml{} services  definition on experimentation and workflows. For example, \textit{Expos\'{e}} \cite{vanschoren2010expose} allows you to focus your work on the experimentation workflow \cite{marozzo2016workflow}. With \textit{OntoDM} \cite{panov2008ontodm} and using \textit{BFO} (Basic Formal Ontology), it is aimed to create a framework for data mining experimentation and replication. In \textit{MEXcore} \cite{esteves2015mex} and \textit{MEXalgo} \cite{esteves2015interoperable} a complete specification of the process of description of experimentation in \dm{} problems is made. In addition, together with \textit{MEXperf} \cite{esteves2015mex}, which adds performance measures to the experimentation, the definition of this type of schema is completed. Another approach similar to \textit{MEX} is \textit{ML-Schema} \cite{w3mlschema}, which attempts to establish a standard schema for algorithms, data and experimentation. \textit{ML-Schema} consider terms like Task, Algorithm, Implementation, Hyper-Parameter, Data, Features or Models.

The most recent proposals try to unify different schemata and vocabularies previously developed following the \linkeddata{} \cite{berners2001semantic} guidelines. With \linkeddata{}, you can reuse vocabularies, schemata and concepts. This significantly enriches the definition of the schema, allowing you to create the model definition based on other existing schemata and vocabularies. \linkedusdl{}, \textit{MEX[core,algo,perf]}, \textit{ML-Schema}, \textit{OntoDM} or \textit{Expos\'{e}} among others, can be considered when creating a workflow for \dmml{} service definition using \linkeddata{}. These proposals provide the definition of consistency in the main area of the service of \dm{} together with the \linkeddata{} properties, enabling the inclusion of other externals schemata that complement the key aspects of a \cc{} service fully defined. 

With this review of state-of-the-art research material, a part of the range of services definition proposals have been studied in order to describe \cc{} services from different points of view. In this proposal we seek to bring together different proposals and some new ones in order to create a broad definition of \dm{} services.

\section{\dmccschema{}: \datamining{} services with \linkeddata}
\label{sec:dmcc-schema}

\sw{} applied to the definition of \cc{} services, allow  tasks such as negotiation, composition and invocation with a high degree of automation. This automation, based on \linkeddata{}  is fundamental in \cc{} because it allows services to be discovered and explored for consumption by other entities using the full potential of \textit{\rdf{}} and \sparql{}. \linkeddata{} \cite{heath2011linked} offers a growing body of reusable schemata and vocabularies for the definition of \cc{} services of any kind \cite{vandenbussche2017linked}. 



In this article we propose \dmccschema{}, a schema and a set of vocabularies which has been designed to address the problem of describing and defining \dmml{} services in \cc{}. Not only it focuses on solving the specific problem of modelling, with the definition of workflow and algorithms, but it also includes the main aspects of a \cc{} service. \dmccschema{} can be considered as a \linkeddata{} proposal for \dmml{} services. Existing \linkeddata{} vocabularies have been integrated into \dmccschema{} and new vocabularies have been created \textit{ad-hoc}  to cover certain aspects that are not implemented by other external schemata. Vocabularies have been re-used following \linkeddata{}  recommendations, filling important parts such as the definition of experiments and algorithms, as well as the interaction or authentication that were already defined in other vocabularies. A service offered from a \cc{} provider, must have the following aspects for its complete specification:


\begin{itemize}
\item \textbf{Authentication}. The service or services require authenticated access.
\item \textbf{Catalogue}. The provider has a catalogue of services ready to be discovered and used.
\item \textbf{Entities}. Services interact between entities offering or consuming services.
\item \textbf{Interaction}. Access points and interfaces for services consumption for users and entities.
\item \textbf{Prices}. The services offered have a cost.
\item \textbf{\sla{}/\slo{}}. The services have service level agreements and more issues.
\end{itemize}

There is no standardization about what elements a service in \cc{} should have for its complete definition, but according to \textit{NIST} \footnote{National Institute of Standards and Technology, www.nist.gov}, it must meet aspects such as self-service (discovery), measurement (prices, \sla{}), among others. For the development of \dmccschema{}, a comprehensive study of the features and services available on the \dmml{} platforms on \cc{} has been carried out. Table \ref{dmservices} contains information about the \dm{} services analyzed from some \cc{} providers; leading providers  such as \textit{Google}, \textit{Amazon}, \textit{Microsoft Azure}, \textit{IBM} and \textit{Algorithmia}, for whom, \sla{}, pricing for the different variants and conditions,  service catalogue (\dmml{} algorithms), methods of interaction with the service, and authentication have been studied.

\begin{table}
\centering
\caption{\cc{} Data Mining services analyzed.}
\label{dmservices}
\begin{tabular}{@{}ll@{}}
\toprule
Provider  & Service name \\ \midrule
  
 \textbf{Google} & Cloud Machine Learning Engine                                                    \\
\textbf{Amazon} & \begin{tabular}[l]{@{}l@{}}Amazon SageMaker, Amazon Machine Learning\end{tabular}                            \\
\textbf{IBM}    & Watson Machine Learning, Data Science                                                    \\
\textbf{Microsoft Azure}  & \begin{tabular}[l]{@{}l@{}} Machine Learning Studio\end{tabular}                 \\
\textbf{Algorithmia}  & \begin{tabular}[l]{@{}l@{}} Algorithms bundles\end{tabular}                 \\  \bottomrule
 
\end{tabular}
\end{table}

\begin{table}[]
\centering
\caption{Features included with  compared with other. }
\label{tab:schemata}
\begin{tabular}{lccccc}
\toprule
 & dmcc & mls\cite{mlschema} & Expos\`e\cite{vanschoren2010expose}  & MEX\cite{esteves2015mex} &L-USDL\cite{bizer2009linked} \\ \midrule
Task & $\bullet$ & $\bullet$ & $\bullet$ & $\bullet$ & - \\ \hline
Impl. & $\bullet$ & $\bullet$ & $\bullet$ & - & - \\\hline
Data & $\bullet$ & $\bullet$ & - & $\bullet$ & - \\ \hline
Model & $\bullet$ & $\bullet$ & $\bullet$ & $\bullet$ & $\bullet$ \\\hline
Auth. & $\bullet$ & - & - & - & $\bullet$ \\ \hline
Pricing & $\bullet$ & - & - & - & $\bullet$ \\ \hline
SLA & $\bullet$ & - & - & - & $\bullet$ \\ \hline
Instan. & $\bullet$ & - & - & - & - \\ \bottomrule
\end{tabular}
\end{table}

Our schema has been complemented reusing external vocabularies: the interaction with the service, using \textit{schema.org} \cite{Guha:2015:SES:2857274.2857276}, authentication, with \textit{Web Api Authentication} (\textit{waa}) \cite{maleshkova2010using}, price design, with \textit{GoodRelations}  \cite{hepp2008goodrelations} and experimentation and algorithms with \dmml{}, using the \dmccschema{} vocabulary (\textit{mls}). Table \ref{vocabs} shows the vocabularies reused in the \textit{mls} schema to define each module of the full service. In table \ref{tab:schemata} a comparison between \dmccschema{} and other schemata  and vocabularies is performed. 

\begin{table}
\centering
\caption{Vocabularies reused by \dmccschema{}.}
\label{vocabs}
\begin{tabular}{@{}ll@{}}
\toprule
\textbf{Module}  & \textbf{Reused Vocabularies } \\ \midrule
\sla              & gr, schema, ccsla     \\
Pricing          & gr, schema, ccpricing, ccinstance, ccregion            \\
Authentication   & waa                   \\
Interaction      & schema                \\
Data Mining      & mls, ccdm                   \\
Service Provider & gr, schema                    \\ \bottomrule
\end{tabular}
\end{table}

\begin{figure*}
  \centering
   {\epsfig{file = 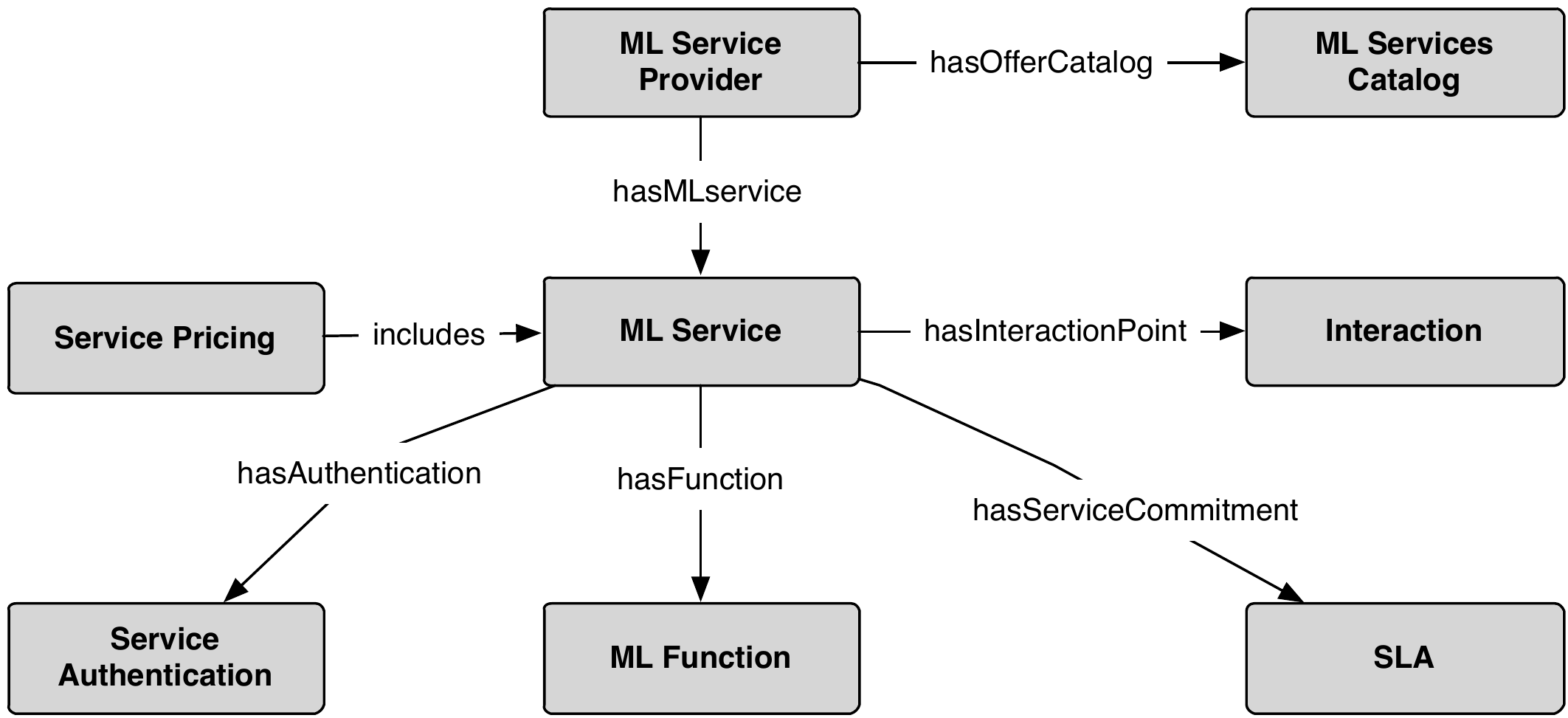, width = 12cm}}
  \caption{Main classes, and relations for \dmccschema{} (\textit{dmcc}).}
  \label{fig:schema}
 \end{figure*}

The high level diagram of \dmccschema{} entities can be seen in the figure \ref{fig:schema}. The detailed definition of each entity is developed in the following subsections.

\dmccschema{} allows to structure the data and information of the services in a coherent way, normalizing the properties and the main concepts of the CC from the different DM service providers. Regarding the benefits of using \dmccschema{} over other schemata, a) offers a flexible data structure that can integrate the properties of existing DM services, b) unifies different schemata  into one and offers friction less integration of the different schemata included, and c) comprises an all-in-one solution for the definition of CC services for DM. With \dmccschema{} we  have tried to emphasize  the independence of the \cc{} platform, that is, \dmccschema{} does not define aspects of service deployment or elements closer to implementation tools. \dmccschema{} definition is at the highest level and only addresses the definition and modelling aspects of the service without going into the details of infrastructure deployment. Below we detail each of the parts that compose our proposal.

\subsection{Authentication}

Nowadays, security in computer systems and in particular in \cc{} platforms is an aspect that must be taken very seriously when developing \cc{} services and applications. In this way, a service that is reliable and robust must also be secure against potential unauthorized access. In the outer layer of security in \cc{} services consumption, authentication that should be considered as a fundamental part of a \cc{} service definition.

\begin{figure}
  \centering
   {\epsfig{file = 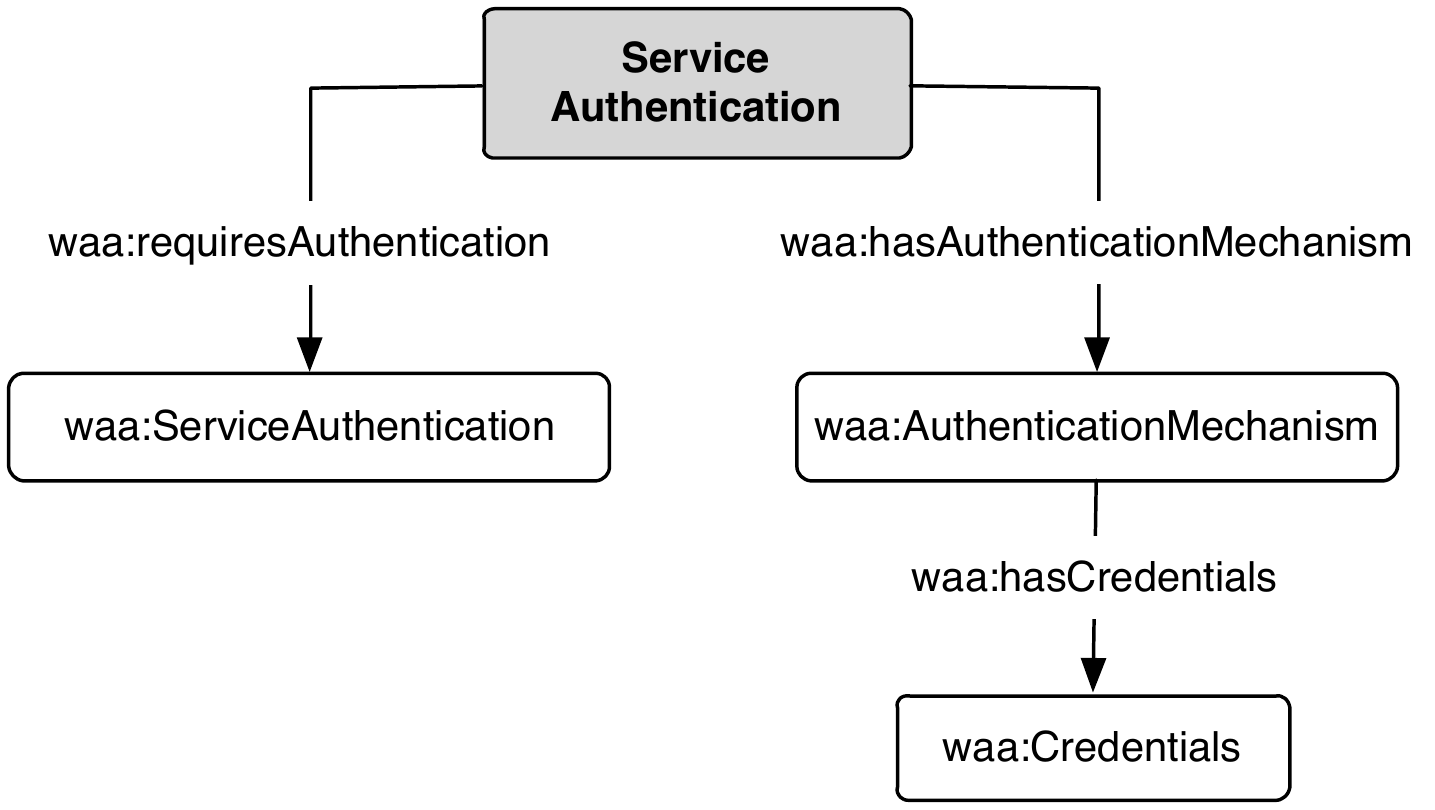 ,width = 9cm}}
  \caption{Authentication schema.}
  \label{fig:auth}
 \end{figure}

Authentication on \cc{} platforms covers a wide range of possibilities. It should be noted that for the vast majority of services, the most commonly used option for managing user access to services are \textit{API Key} or \textit{OAuth} and other mechanisms. \textit{waa:WebApiAuthentication} has been included for authentication modelling. This schema allows to model many of the authentication systems available. In figure \ref{fig:auth}, the model for the definition of authentication services is depicted, together with the details of the authentication mechanisms.

\subsection{Data Mining service}

\begin{figure*}
  \centering
   {\epsfig{file = 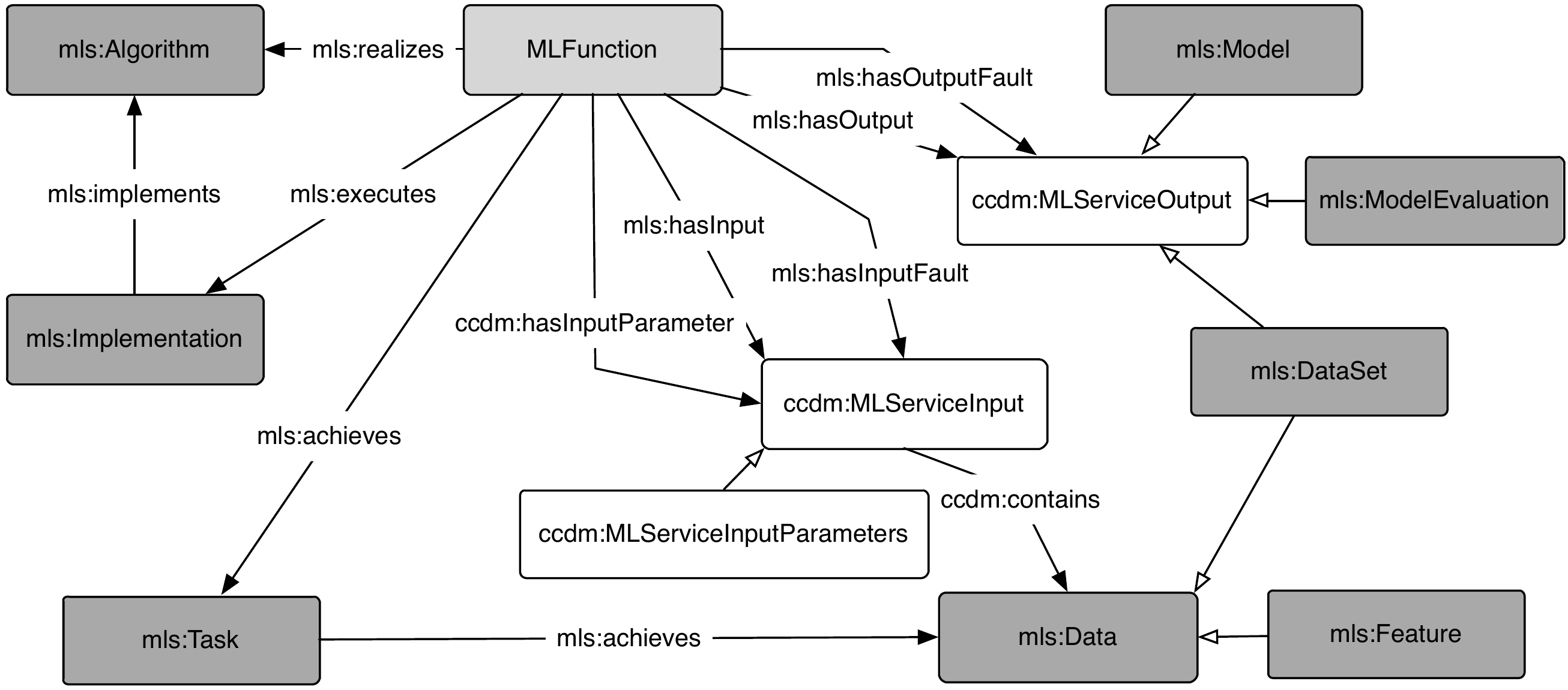, width = 15cm}}
  \caption{\dm{} experimentation, workflow and algorithms execution.}
  \label{fig:mls}
 \end{figure*}

For the main part of the service, where the experimentation and execution of algorithms is specified and modelled, parts of \textit{ML-Schema} (mls) have been reused. \textit{MEXcore, OntoDM, DMOP} or \textit{Expos\'{e}} also provide an adequate abstraction to model the service, but they are more complex and their vocabulary is more extensive. \textit{ML-Schema} has been designed to simplify the modelling of \dmml{}  experiments and bring them into line with that which is offered by \cc{} providers. We have extended \textit{ML-Schema} by adapting its model to a specific one and inheriting all its features (figure \ref{fig:mls}). 

The following vocabulary components are highlighted (\textit{ccdm} is the name of the schema used): 

\begin{itemize}
\item \texttt{ccdm:MLFunction} Set the operations, function or algorithm to be executed. For example \texttt{Random Forest} or \texttt{KNN}. 
\item \texttt{ccdm:MLServiceOutput} The output of the algorithm or the execution. Here the output of the experiment is modelled as Model, Model Evaluation or Data.
\item \texttt{ccdm:MLServiceInput} The algorithm input, which corresponds to the setting of the algorithm implementation. Here, you can describe the model the data entry of the experiment, such as the data set and parameters (\texttt{ccdm:MLServiceInputParameters}) of the algorithm executed.
\item \texttt{mls:Model} Contains information specific to the model that has been generated from the run.
\item \texttt{mls:ModelEvaluation} Provides the performance measurements of the model.
\item \texttt{mls:Data} They contain the information of complete tables or only attributes (table column), only instances (row), or only a single value.
\item \texttt{mls:Task} It is a part of the experiment that needs to be performed in the \dmml{} process.
\end{itemize}

\subsection{Interaction}

The interaction with \dmml{} services is generally done through a \restfulapi{}. This API provides the basic functionality of interaction with the service consumer, who must be previously authenticated to use the services identified in this way. For the interaction the \textit{Action} entity of the vocabulary \textit{schema} was used. With this definition, the service entry points, methods and interaction variables are fully specified for all services specified by the API.

\subsection{\sla{}: Software Level Agreement}

The trading of \cc{} services sets up a series of contractual agreements between the stakeholders involved in the services. Both the provider and the consumer of the service must agree on service terms. The service level agreements define technical criteria, relating to availability, response time or error recovery, among others. The \slo{} are specific measurable features of the \sla{} such as availability, throughput, frequency, response time, or quality of service. In addition with the \slo{}, it needs to contemplate actions when such agreements cannot be achieved where in this case compensation is offered. 

The \sla{} studied for the \dmml{} service environment are established by terms and definitions of the agreements (\cite{ambulkar2012data}, \cite{li2013adaptive}, \cite{zheng2013service} ). These terms may have certain conditions associated with them which, in the case of violation, involve a compensation for the guarantee service.

In general terms, \sla{} for \cc{} services is given by a term of the agreements that contains one or more definitions similar to \textit{Monthly Up-time Percentage} (MUP), which specifies the maximum available minutes less downtime divided by maximum available minutes in a billing month. For this term, a metric or interval is established over which a compensation is applied in the case that the agreement term  is not satisfied. In this context a schema named \textit{ccsla}\footnote{Available at  \url{http://lov.okfn.org/dataset/lov/vocabs/ccsla}} has been created for the definition of all the components of an \sla{}. In Table \ref{termssla} an example with a pair of MUP intervals and compensation related  is shown.

\begin{table}
\centering
\caption{Example of a MUP term and compensations related.}
\label{termssla}
\begin{tabular}{@{}ll@{}}
\toprule
\textbf{Monthly Uptime Percentage (MUP)}  & \textbf{Compensation } \\ \midrule
	$[$0.00\%, 99.00\%$]$          & 25\%           \\
    $[$99.00\%, 99.99\%$]$        & 10\%      \\  \bottomrule
\end{tabular}
\end{table}

\begin{figure}
  \centering
   {\epsfig{file = 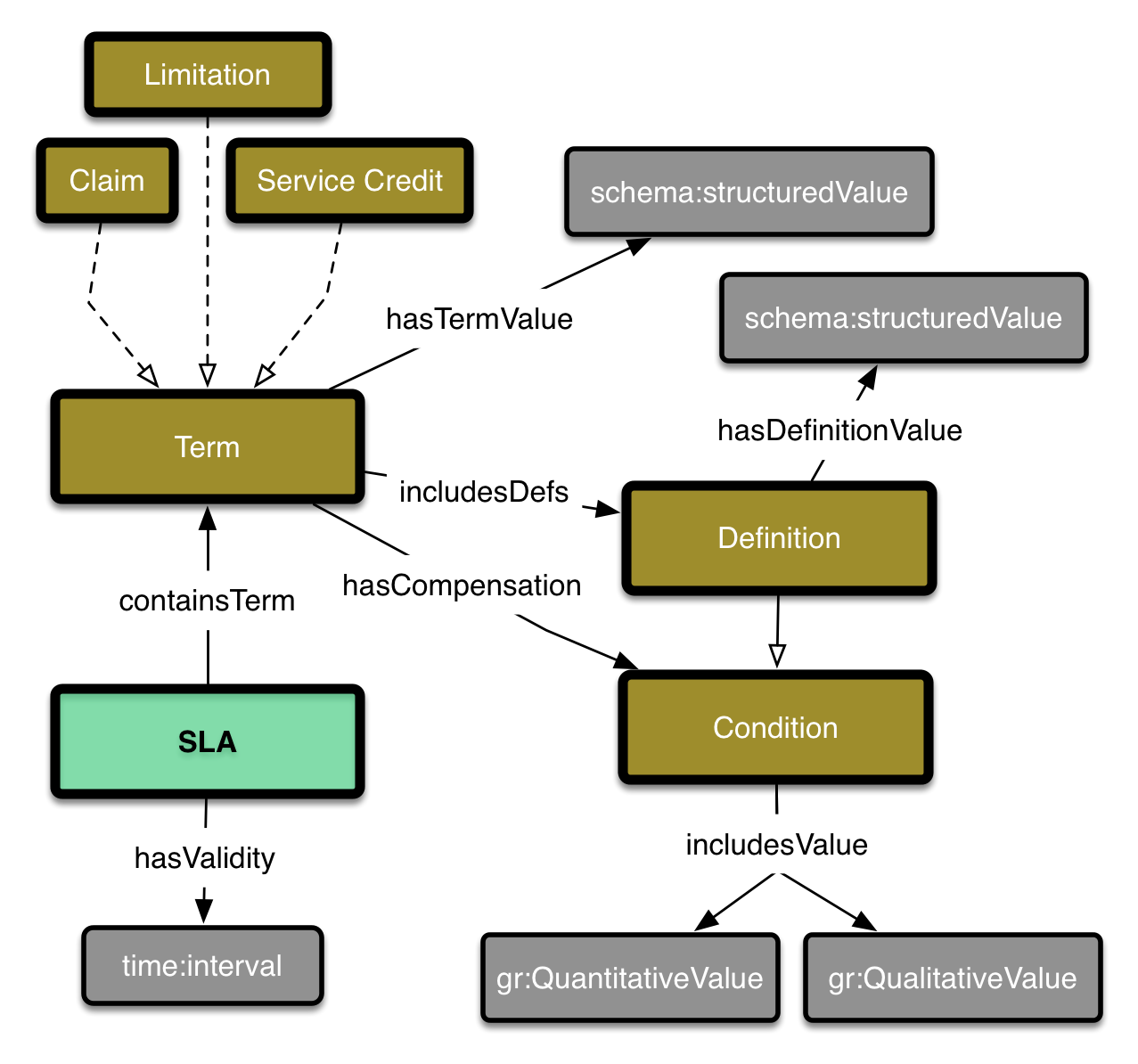 ,width = 12cm}}
  \caption{\sla{} schema for \cc{} services (\textit{ccsla}).}
  \label{fig:sla}
 \end{figure}

In figure \ref{fig:sla} the complete schema of the \sla{} modelling that has been designed is illustrated.

\subsection{Pricing}

Like any other utility-oriented service, \cc{} services are affected by costs and pricing that vary with the price of the service offered. Following the pay-as-you-go model, the costs of using the service are directly related to the characteristics of the service and the use made of it. The pricing of services in \cc{} is a complex task. Not only there is a price plan for temporary use of resources, but it is also affected by technical aspects, configuration or location of the service. Table \ref{pricestable} collects some of the price modifying aspects of the use of the \dmml{} service.  In the specific segment of \dm{} services in \cc{} there are several elements to consider.

\begin{table}
\centering
\caption{Price components for each \cc{} provider analyzed.}
\label{pricestable}
\begin{tabular}{@{}llll@{}}
\toprule
\textbf{Provider / Service}  & \textbf{Region} & \textbf{Instances} &  \textbf{Storage}  \\ \midrule
  Amazon SageMaker     & y      & y                                                            & y            \\
  Amazon Lambda        & y      & n                                                            & n          \\
Azure Machine Learning             & y      & y                                                            & y            \\
Azure Functions      & y      & n                                                            & n            \\
IBM BlueMix          & y      & y                                                            & n          \\
Google Machine Learning            & y      & y                                                            & n         \\
Algorithmia           & n      & n                                                            & n          \\   \bottomrule
\end{tabular}
\end{table}

\begin{figure}
  \centering
   {\epsfig{file = 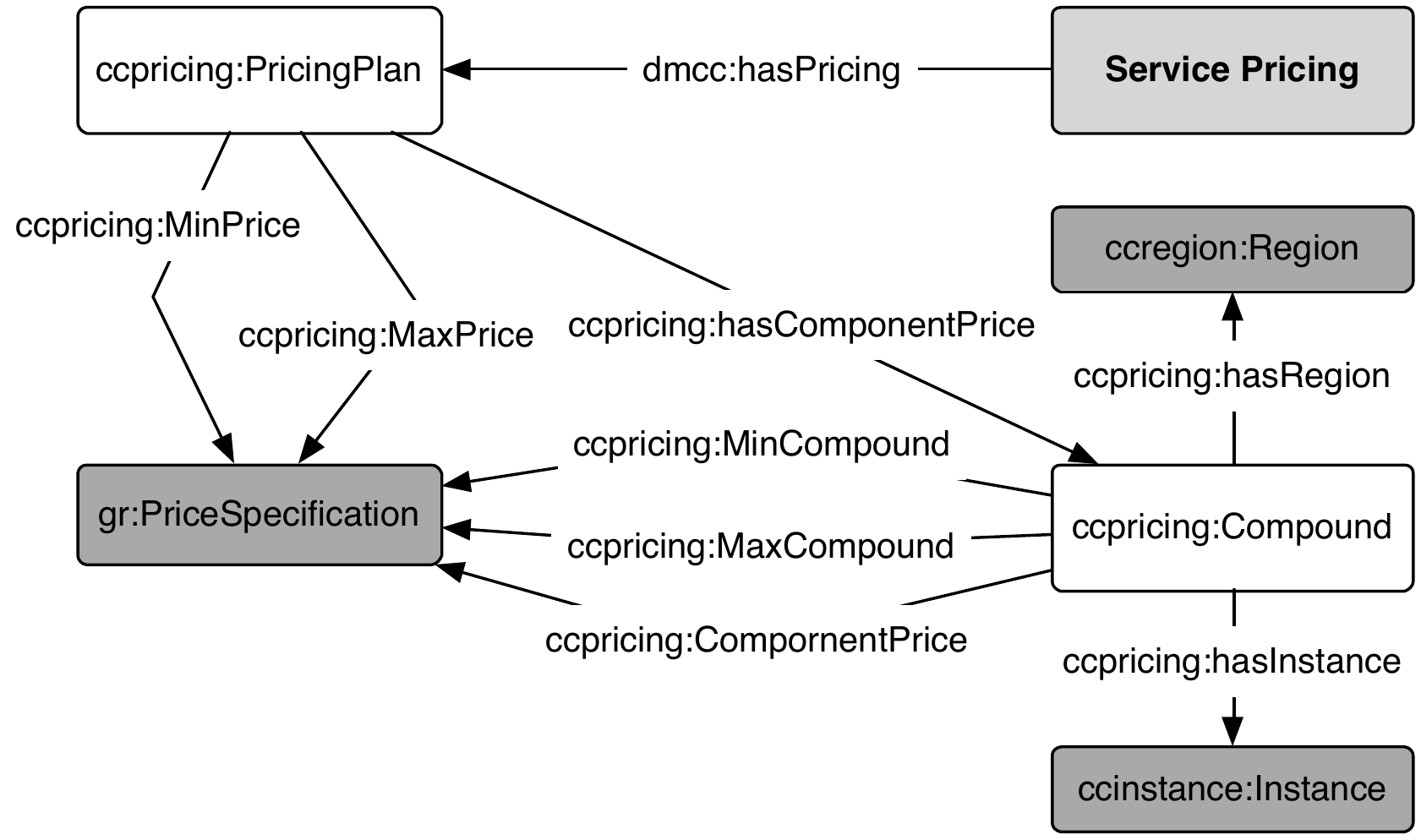 ,width =  12cm}}
  \caption{Pricing schema with \textit{ccpricing}.}
  \label{fig:prices}
 \end{figure}

The following components of the vocabulary stand out from the diagram of figure \ref{fig:prices}: 

\begin{itemize}
\item \texttt{ccpricing:PricingPlan}. It allows to define the attributes of the price plan and their details. For instance, a free, premium plan, or similar. It must match the attributes of \textit{MinPrice} and \textit{MaxPrice}. 
\item \texttt{gr:PriceSpecification}. It provides all the tools to define the prices of the services at the highest level of detail. Comes from \textit{GoodRelations} (\textit{gr}) schema.
\item \texttt{ccpricing:Compound}. Components of the price of the service. Services are charged according to values that are related to certain attributes of the service configuration. In compound you can add aspects such as type of instances, cost of the region, among others.
\item \texttt{ccinstances:Instance}. Provides the specific vocabulary for the definition of price attributes related to the instance or instances where the \dm{} service is executed.
\item \texttt{ccregions:Region}. It allows to use of locations and regions schema, offered by providers. This is an additional value to the costs of the service, as part of the price.
\end{itemize}

For the modelling of this part, three vocabularies supplementing the definition of prices, have been developed which were not available. On the one side  \textit{ccinstances}\footnote{\url{http://lov.okfn.org/dataset/lov/vocabs/cci}}, which contains everything necessary for the definition of instances (\textit{CPU}, number of \textit{CPU} cores, model, \textit{RAM} or \textit{HD}, among others), on the other side \textit{ccregions} \footnote{http://lov.okfn.org/dataset/lov/vocabs/ccr}, which provides the general modelling of the service regions and location  in \cc{} and the last one the schema for the prices modelling \textit{ccpricing} \footnote{http://lov.okfn.org/dataset/lov/vocabs/ccp} .

The complete diagram of the \dmccschema{} is available on the project website \cite{parradmservices}, from which you can explore the relationships, entities, classes, and attributes of each of the modules that comprise the proposal.

\section{Use cases and validation}
\label{sec:usecases}

For reasons of space we will not detail all the data or attributes in depth and will only consider what is most important for the basic specification of the service and his comprehension.  All detailed information on each component, examples and data sets are available on the project's complementary information website.

In this section we will illustrate how \dmccschema{} can be used effectively to define a specific \dm{} service, as well as additional aspects related to \cc{} such as \sla{}, interaction, pricing or authentication. To do this we will create a service instance for a \rf{} \cite{ho1998random} algorithm for classification. In \rf{} service implementation, we will use the default function model and parameters of this algorithm in \rlang{} \cite{team2013r} language. We have chosen the parameters format of the \rlang{} algorithms, due to the great growth that this programming language is currently having for data science. In order to instantiating of the service we will use vocabularies such \textit{dcterms} (\texttt{dc:}) \cite{weibel1998dublin}, \textit{GoodRelations} (\texttt{gr:}) or \textit{schema.org} (\texttt{s:}).  In these cases of use we will name \dmccschema{} as \texttt{dmcc} namespace.

The entry point for defining the service is the instantiating of the \texttt{dmcc:ServiceProvider} entity, including its data, using the class \texttt{dmcc:MLServiceProvider}. We will specify an example of \dm{} service hosted on our \cc{} platform \textit{dicits.ugr.es}.

\begin{scriptsize}
\begin{lstlisting}[basicstyle=\ttfamily\small,showspaces=false,numbers=left,caption=New DM service definition.,label=code1, xleftmargin=\parindent,captionpos=b]
_:MLProvider a dmcc:MLServiceProvider;
 rdfs:label "ML Provider"@en ;
 dc:description 
   "DICITS ML SP"@en ;
 gr:name "DITICS ML Provider";
 gr:legalName "U. of Granada";
 gr:hasNAICS "541519";
 s:url <http://www.dicits.ugr.es>;
 s:serviceLocation 
  [ a s:PostalAddress;
    s:addressCountry "ES";
    s:addressLocality "Granada";
  ] ;
  s:contactPoint 
  [
   a s:ContactPoint;
   s:contactType "Costumer Service";
   s:availableLanguage [ 
     a s:Language;
     s:name "English";];
   s:email "ml@dicits.ugr.es";
  ];		
 dmcc:hasMLService 
          _:MLServiceDicitsRF,
          _:MLServiceDicitsKMeans;
 dmcc:hasOfferCatalog 
          _:MLServiceDicitsCatalog;
.	
\end{lstlisting}
\end{scriptsize}

The code above shows the specific details of the service provider, such as legal aspects, location or contact information. The service provider \texttt{dmcc:MLServiceProvider} has the property \texttt{dmcc:hasMLService}. With this property two services such as \texttt{\_:MLServiceDicitsRF} and \texttt{\_:MLServiceDicitsKMeans} (listing \ref{code1} line 24-25) are defined.

Listing \ref{code2} shows the instantiating of each of the entities for \texttt{\_:MLServiceDicitsRF}. \texttt{\_:MLServiceDicitsRF} allows defining all the service entities such as \sla{} (listing \ref{code2} line 8),  algorithm (listing \ref{code2} line 10), prices (listing \ref{code2} line 15), interaction points (listing \ref{code2} line 6) and  authentication (listing \ref{code2} line 12).

\begin{scriptsize}
\begin{lstlisting}[basicstyle=\ttfamily\small,showspaces=false,numbers=left,caption=Components of the DM service ,label=code2, xleftmargin=\parindent,captionpos=b]
_:MLServiceDicitsRF a dmcc:MLService;
 rdfs:label 
    "ML Service dicits.ugr.es"@en ;
 dc:description 
    "DICITS ML Service"@en ;
 dmcc:hasInteractionPoint 
    _:MLServiceInteraction;
 dmcc:hasServiceCommitment 
    _:MLServiceSLA;
 dmcc:hasFunction 
    _:MLServiceFunction;
 dmcc:hasAuthentication 
    _:MLServiceAuth;
 dmcc:hasPricingPlan 
   _:MLServicePricing;
.
\end{lstlisting}
\end{scriptsize}

The interaction with the service  is performed using the \texttt{dmcc:Interaction} class that includes the property \texttt{dmcc:hasEntryPoint} that allows to define an \texttt{Action}  on a resource or object. In this case we use the vocabulary \texttt{schema.org}  to set the method of access to the \rf{} service, for which we specify that the service will be consumed through a \restfulapi{}, with HTTP  \texttt{POST} access method, the URL template \texttt{http://dicits.ugr.es/ml/rf/}, for instance and its parameters.



The selected service requires authentication. An \apikey{} authentication has been chosen. The instantiating of service authentication is defined requesting authentication with \texttt{waa:requiresAuthentication} and setting up the authentication mechanism \texttt{waa:hasAuthenticationMechanism} to  \texttt{waa:Direct} and as credentials an \apikey{} is also used as shown in listing \ref{code4}. 

\begin{scriptsize}
\begin{lstlisting}[basicstyle=\ttfamily\small,showspaces=false,numbers=left,caption=Service authentication.,label=code4, xleftmargin=\parindent,captionpos=b]
_:MLServiceAuth 
     a dmcc:ServiceAuthentication;
 rdfs:label 
   "Service Authentication"@en ;
 dc:description "Service Auth"@en ;	
 waa:requiresAuthentication waa:All;
 waa:hasAuthenticationMechanism 
   [ a waa:Direct ;
     waa:hasInputCredentials 
    [ a waa:APIkey;
     waa:isGroundedIn "key";
    ];
  waa:wayOfSendingInformation 
      waa:ViaURI;
 ]
.
\end{lstlisting}
\end{scriptsize}


For the definition of the \sla{}, in our example service we have taken the some providers as \amazon{} or \azure{} where they identify a term \mup{}. For the cases where this occurs \sla{} define two ranges: less than \textit{99.00} is compensated by \textit{30} credits  (in service usage) and the range of \textit{99.00} to \textit{99.99} is compensated with \textit{10} credits. To model the \sla{} we use \texttt{ccsla:SLA} together with the property \texttt{ccsla:cointainsTerm} \_:SLATermMUP\_A; in which we define the specific terms.



To define the range of the term is used in \texttt{\_:SLADefinition\_A},  as shown in  listing \ref{code6}.

\begin{scriptsize}
\begin{lstlisting}[basicstyle=\ttfamily\small,showspaces=false,numbers=left,caption=SLA term definition.,label=code6, xleftmargin=\parindent,captionpos=b]
_:SLADefinition_A a ccsla:Definition;
 ccsla:hasDefinitionValue [ 
 a s:structuredValue;
 s:value [ 
  a s:QuantitativeValue;
  s:maxValue 99.99;
  s:minValue 99.00;
  s:unitText "Percentaje";
  ];
 ];
.
\end{lstlisting}
\end{scriptsize}

The schema for SLA (\textit{ccsla}) and other examples with the \sla{} for \textit{Amazon Web Services} and \textit{Microsoft Azure} can be accessed from the website of the project \cite{parradmservices}.

For the definition of the economic cost  of the service we have considered two variants for the example. The first is for free service use, limited to \textit{250} hours of execution of algorithms within an instance (Virtual Machine) with a CPU model \textit{Intel i7},  64 GB of RAM and one region. The second pricing model where you charge what you consume for the service in \textit{USD/h.}, for an instance and one region.



We can define multiple pricing plans, for this example a free plan is  specified with      \texttt{\_:ComponentsPricePlanFree}. The price modelling is done with our proposal using the definition of prices provided by \textit{ccpricing}.


For each price plan we take into account the variables and features that affect the price. These are: region, instance type and other components using  \texttt{\_:ComponentsPricePlanFree}.


For example, to define features of the type of instance used in the free plan, we use \texttt{ccinstance:Instance;} and a few attributes like \ram{} or \cpu{} as seen in listing   \ref{code86}. More examples for the schema \textit{ccinstances}, including  a small dataset of  Amazon instances \cite{parradmservices} (types \textit{T1} and \textit{M5}) are available in the web site of the complementary information of the paper.

\begin{scriptsize}
\begin{lstlisting}[basicstyle=\ttfamily\small,showspaces=false,numbers=left,caption=Pricing and instances.,label=code86, xleftmargin=\parindent,captionpos=b]
_:InstanceFree 
    a ccinstances:Instance;
   ccinstances:hasRAM [ 
     a ccinstances:ram;
        s:value "64"
        s:unitCode "E34";
   ] ;
   ccinstances:hasCPU [ 
     a ccinstances:cpu;
      ccinstances:cpu_model 
        "Intel i7";      
   ] ;
.	
\end{lstlisting}
\end{scriptsize}

In order to define the \texttt{\_:MaxUsageFree}, we need to determine the free access plan to the service and the limitation of compute hours to 250.  For this purpose we use \texttt{gr:PriceSpecification} and \texttt{gr:Offering} classes as shown in listing  \ref{code9}.

\begin{scriptsize}
\begin{lstlisting}[basicstyle=\ttfamily\small,showspaces=false,numbers=left,caption=Price specification for the free plan.,label=code9, xleftmargin=\parindent,captionpos=b]
_:MaxUsageFree a 
   gr:PriceSpecification, 
   gr:Offering;
 gr:max 0.00;
 gr:priceCurrency "USD";
 gr:includesObject [
  a gr:TypeAndQualityNode;
  gr:amountOfThisGood "250";
  gr:hasUnitOfMeasurement "HRS";
 ];
.
\end{lstlisting}
\end{scriptsize}



Additional examples  for the \textit{ccpricing} schema, and a dataset of  \textit{Amazon SageMaker} pricing plans, are available in \cite{parradmservices}.


Finally, we define the service algorithm using \texttt{ccdm:MLFunction} for the definition of a \rf{} function, where we specify the input parameters (data set and hyper-parameters), the output of the algorithm, among others (see listing \ref{code31}).

\begin{scriptsize}
\begin{lstlisting}[basicstyle=\ttfamily\small,showspaces=false,numbers=left,caption=Operations for the algorithm/function.,label=code31, xleftmargin=\parindent,captionpos=b]
_:RandomForest_Function 
  a ccdm:MLFunction ;
   ccdm:hasInputParameters 
     _:RF_InputParameters;
   mls:hasInput 
    _:RF_Input;
   mls:hasOutput 
    _:RF_Output .
\end{lstlisting}
\end{scriptsize}

The  input and output data of the algorithms must be included in the definition of the data mining operation to be performed. The input of data, which can be parameters \texttt{\_:RF\_InputParameters}  or data sets  \texttt{\_:RF\_Input}. Input parameters of the algorithm can be defined with \texttt{dmc:MLServiceInputParameters} and the parameter list \texttt{\_:parameter\_01, [...]}.


Definition of \texttt{ccdm:hasInputParameters \_:RF\_InputParameters} allows you to specify the general input parameters of the algorithm. For example for \rf{} \texttt{dc:title "ntrees"} (number of trees generated), as well as whether \texttt{ccdm:mandatory "false"} is mandatory and its default value, if it exists. Listing \ref{code51} shows the definition of one of the parameters \texttt{parameter\_01}. The other algorithm parameters are defined in the same way.

\begin{scriptsize}
\begin{lstlisting}[basicstyle=\ttfamily\small,showspaces=false,numbers=left,caption=Example of parameter and features.,label=code51, xleftmargin=\parindent,captionpos=b]
_:parameter_01 
  a ccdm:MLServiceInputParameter ;
    ccdm:defaultvalue "100" ;
    ccdm:mandatory "false" ;
    dc:description 
      "Number of trees" ;
    dc:title "ntrees" .
\end{lstlisting}
\end{scriptsize}

An \texttt{mls:Model} model and an evaluation of the \texttt{mls:ModelEvaluation} model have been considered for specifying the results of the \rf{} service execution in \texttt{mls:hasOutput \_:RF\_Output}. Model evaluation is the specific results if the algorithm returns  a value or set of values. When the service algorithm is pre-processing the result is a data set. For the model you have to define for instance whether the results are PMML \cite{guazzelli2009pmml} with \texttt{\_:RF\_Model a dmc:PMML\_Model} as shown in listing \ref{code61}.

\begin{scriptsize}
\begin{lstlisting}[basicstyle=\ttfamily\small,showspaces=false,numbers=left,caption=PMML Model and storage of the service output.,label=code61, xleftmargin=\parindent,captionpos=b]
_:KMeans_Model 
  a ccdm:PMML_Model ;
    ccdm:storagebucket 
      <dicits://models/> ;
    dc:description 
      "PMML model" ;
    dc:title "PMML Model" .
\end{lstlisting}
\end{scriptsize}



Other services implemented on \dmccschema{} as examples, such as \textit{NaiveBayes}, \textit{LinearRegression}, \textit{SVM} or  \textit{Optics}, are available in the supplementary website of the paper \cite{parradmservices}. 


In order to validate the scheme, a dataset has been created containing the \dmccschema{}  description of several \cc{} providers and their respective algorithms as services. For each provider, the specific data of the service has also been described, such as regions,  instances and economic cost of each one of the variants of the consumption of the services. To confirm the validity of \dmccschema{}, multiple queries with \sparql{} have been carried out to extract information from the dataset, such as to know the providers that offer a specific \dm{} service or to obtain the best price to run a \rf{} algorithm. Dataset and query results can be checked from the project site \cite{parradmservices}.

\section{Conclusion}
\label{sec:conclusions}

In this article we have presented \dmccschema{}, a simple and direct schema for the description and definition of \dm{} services in \cc{}. Our proposal tries to gather, on the one hand, everything related to the definition of the experimentation, workflow and algorithms and on the other hand, all the other aspects that compose a complete CC service. Our schema has been built on the basis of \sw{}, using an ontology language to implement it and following the \linkeddata{} directives regarding the re-use of other schemata, which perfectly enrich the service modelling that has been designed.

\dmccschema{} is presented as a light-weight tool for services modelling that allows the creation of a complete \dm{} service that includes all providers of the \cc{} platforms, adapting in a flexible way to the differences of definition and description of services of the most well-known providers.

The example of use shown, illustrates the effortless definition of a service whose objective is to execute a simple \rf{} algorithm, and indicating other aspects related to the \cc{} service itself.

One of the advantages of using \dmccschema{} is that it abstracts differences between heterogeneous \cc{} providers for \dm{}  services in order to have a single and unique specification that can bring together different services specifications. In this way the differences between the definitions are balanced, allowing to \dmccschema{} to be used as an integral part of a \cc{} Services Broker, storing such services from different providers.

Finally, it is important to highlight that \dmccschema{} is being used successfully within a computing and workflow platform for \dm{}, called \occml{}. As part of the platform, \dmccschema{} is used to define and describe complete \dm{} services, allowing a high degree of flexibility and portability.

\section*{Acknowledgment}

M. Parra-Royon holds a "Excelencia" scholarship from the Regional Government of Andaluc\'ia (Spain). This work was supported by the Research Projects \textit{P12-TIC-2958}  and \textit{TIN2016-81113-R} (Ministry of Economy, Industry and Competitiveness - Government of Spain).




%

\bibliographystyle{IEEEtran}
\bibliography{IEEEabrv,references}

\end{document}